\documentclass[final,3p,times]{elsarticle}

\usepackage{soul}
\usepackage[usenames,dvipsnames]{xcolor}

\usepackage{lineno,hyperref}
\usepackage{amsthm}
\usepackage{amssymb}
\usepackage{amsmath}
\usepackage{multirow}
\usepackage{lineno}
\journal{Information Sciences}
\begin{document}

\begin{frontmatter}

%\title{Recommending investors for new startups through diffusion on tripartite networks}
\title{Recommending investors for new startups by integrating network diffusion and investors' domain preference}

\author{Shuqi Xu\fnref{label1}}
\author{Qianming Zhang\fnref{label2}}
\author{Linyuan L\"u\corref{cor1}\fnref{label1,label3}} 
\author{Manuel Sebastian Mariani\fnref{label1,label4}\corref{}}

\address[label1]{Institute of Fundamental and Frontier Science, University of Electronic Science and Technology of China, Chengdu, China}
\address[label2]{CompleX Lab, Big Data Research Center, University of Electronic Science and Technology of China, Chengdu, China}
\address[label3]{Alibaba Research Center for Complexity Sciences, Hangzhou Normal University, Hangzhou, China}
\address[label4]{URPP Social Networks, Universit$\ddot{a}$t Z$\ddot{u}$rich, Z$\ddot{u}$rich, Switzerland}
\cortext[cor1]{Email address: linyuan.lv@gmail.com}

\begin{abstract}
Over the past decade, many startups have sprung up, which create a huge demand for financial support from venture investors. However, due to the information asymmetry between investors and companies, the financing process is usually challenging and time-consuming, especially for the startups that have not yet obtained any investment. Because of this, effective data-driven techniques to automatically match startups with potentially relevant investors would be highly desirable. 
Here, we analyze $34,469$ valid investment events collected from \emph{www.itjuzi.com} and consider the cold-start problem of recommending investors for new startups.
We address this problem by constructing different tripartite network representations of the data where nodes represent investors, companies, and companies' domains. 
First, we find that investors have strong domain preferences when investing, which motivates us to introduce virtual links between investors and investment domains in the tripartite network construction. 
Our analysis of the recommendation performance of diffusion-based algorithms applied to various network representations indicates that prospective investors for new startups are effectively revealed by integrating network diffusion processes with investors' domain preference.
\end{abstract}


\end{frontmatter}

%\linenumbers
\section{\label{sec:Introduction}Introduction}
The rapid development of the Internet brings the information overload problem: people usually receive too much information about a given issue, which greatly impairs the efficiency of their decision-making process~\cite{eppler2004concept}. 
Automated information-filtering tools such as ranking algorithms~\cite{liao2017ranking} and recommender systems~\cite{lu2012recommender} provide us with effective solutions to this problem. In particular, recommender systems can exploit data on users and their past preferences to predict their possible future interests, thus have been widely applied by various online platforms, including e-commerce websites~\cite{linden2003amazon} and online social networks~\cite{tang2013social}. Previous works~\cite{schafer2001commerce} have shown that capable recommender systems can largely increase not only economic benefits but also customer loyalty.

In several financial investment scenarios, including stock investment~\cite{geva2014empirical}, real estates investment~\cite{ginevivcius2011recommender}, crowdfunding project~\cite{an2014recommending}, and portfolio management~\cite{paranjape2013stock}, recommender systems have received increasing attention~\cite{zibriczky2016recommender}. 
Less attention has been devoted to the design and applications of recommendation systems in the domain of venture investment, an emerging investment type that aims at offering seed funding to startup companies.
Extant studies on the topic~\cite{stone2013empirical,zhao2015risk} have focused on helping VC firms to find investee companies, while effective methods to support new startups in their search for investors have not yet gained attention from scholars. For a startup, early-stage fundings constitute key support, yet obtaining them is challenging~\cite{salamzadeh2015startup}. Finding a suitable investor who is interested in the startup's business scope usually requires long-term research, which is especially difficult for new startups due to their inexperience~\cite{davila2003venture}. For this reason, an investor-filtering system aimed at new startups can be extremely beneficial. 

Our main goal is to fill this gap by designing and validating recommendation system techniques to identify suitable investors for new startups. To this end, we analyze 34,469 investment events collected from \url{www.itjuzi.com}~\cite{itjuzi}. As we focus on startups that received no previous investments in the past, widely-studied recommendation techniques based on bipartite networks~\cite{zhou2007bipartite} are not applicable here since the new startups with no previous investors turn out to be isolated nodes in the investor-company bipartite network, making them unreachable by diffusion processes. Our problem can be classified as a \emph{cold-start problem}~\cite{schein2002Methods}: When a new actor enters the system, there is insufficient past information to provide him/her with a recommendation~\cite{lu2012recommender}. 

To overcome the cold-start problem, we resort to tagging techniques~\cite{golder2006usage}.On the \url{www.itjuzi.com} platform, each startup is required to provide several tags that define its business scope. We found that the investors have a strong preference toward a small number of tags -- in particular, in the majority of cases, an investor tends to invest in startups that feature her favorite tag.
Hence, motivated by this finding and the key role played by the industry field in investment decision making~\cite{dushnitsky2005firms,mason2004investors}, we use tags as a key piece of information to generate recommendations. 

By leveraging startups' tag information, we construct three different tripartite network representations of the investment system. The most natural tripartite representation is one where investors are connected to the startups they invested in, and startups are connected to their self-reported tags -- we refer to this representation as the tag-company-investor (TCI) representation. This representation is natural because it is directly based on the collected data (companies' self-reported tags and investor-company investment events), and it is commonly used in the existing studies on online social networks~\cite{shang2010collaborative,zhang2010solving,zhang2010personalized}. This kind of tripartite networks only considers the connections that are already in the data, and the recommendation is made by applying physical diffusion processes such as probabilistic spreading~\cite{zhou2007bipartite} and heat transfer~\cite{zhang2007heat} (see details in Section 4.1) to the available network.
However, we find that if our goal is to produce recommendation of investors for new startups, the natural TCI representation is suboptimal. The reason is that a new startup has not direct connections with investors, which implies that in the TCI representation, relatively long network paths are needed for a diffusion process to travel from a target startup to a prospective investor. Moreover, our empirical analysis indicates that investors tend to concentrate their investments toward startups with a small number of preferred tags. This property suggests that for our problem, investor-tag connections are potentially more informative than investor-company connections. 

Motivated by these observations, we introduce virtual links between investors and tags, and construct two new tripartite network representations: the company-tag-investor (CTI) and the company-investor-tag (CIT) network representations. On each of the three tripartite network representations considered here (TCI, CTI, and CIT), we apply diffusion-based recommendation algorithms. We find that diffusion algorithms based on networks with virtual links (i.e., the company-tag-investor and company-investor-tag networks) achieve substantially better recommendation accuracy compared to those based on the natural tag-company-investor representation. 

Our main focus on diffusion-based techniques is motivated by existing works~\cite{lu2012recommender,yu2016network} that indicate that this class of techniques is well-suited to scenarios where the input data are binary (without ratings) and only limited information about the target items or users is available. At the same time, we do not narrow our focus to this class of algorithms, alternative techniques from existing studies, including neighbor-based collaborative filtering~\cite{sarwar2001item} and matrix factorization based on Bayesian personalized ranking~\cite{gantner2010learning} (see Section~4.2) are also included in the analysis. The proposed diffusion-based algorithms turn out to outperform these existing algorithms that are not based on a tripartite network representation.   

The main contribution of this paper is twofold. First, it brings network-based recommendation techniques to the problem of finding relevant investors for newcomer startups. Second, to address the problem, it introduces virtual links between investors and tags to build the network representation. The possibility to introduce these links has not been exploited by previous studies on recommendation algorithms based on tripartite networks~\cite{shang2010collaborative,zhang2010solving,zhang2010personalized}, yet the virtual links turn out to be vital to achieving a good recommendation performance. The new representation achieves indeed improved performance over recommendation techniques based on the straightforward TCI representation and existing collaborative filtering~\cite{sarwar2001item} and matrix factorization algorithms~\cite{gantner2010learning}.
Our findings reveal a powerful method for startups to find their first investor.

The paper is structured as follows: In Section 2, we review related works. In Section 3, we describe the dataset, analyze the tag preference of investors, and propose the construction of three tripartite network representations. In Section 4, we introduce the recommendation algorithms employed in the tripartite networks, the baseline algorithms, and the evaluation metrics. In Section 5, we present our results and discuss them. Finally, Section 6 concludes the paper and points out several open research directions.

\section{Related work}

\subsection{Recommendation methods} 

Collaborative filtering (CF) is one of the most widely-used family of recommendation techniques. It can be further grouped into memory-based CF and model-based CF~\cite{su2009survey}. In online systems where users rate items, memory-based CF approaches use users' past rating data to compute the similarity between users or items and produce a prediction for the target user based on the similar users to her or similar items to those she already collected. Model-based CF methods leverage historical data to learn a predictive model. Well-known model-based CF techniques include Bayesian networks~\cite{miyahara2000collaborative}, clustering models~\cite{chee2001rectree}, latent semantic models~\cite{hofmann2004latent}, among others. The most critical component of CF methods is the measurement of the similarity between pairs of users or items, which are known to be vulnerable against data sparsity and cold-start problem~\cite{ahn2008new}.

Another common class of techniques is content-based filtering, which utilizes the description of items and the profile of users' preference to find the items that best match the items previously selected by the user.
The recommendation process includes representing the items' features, learning users' profile, and filtering.
There are also hybrid approaches which combine collaborative with content-based methods or with different variants~\cite{burke2002hybrid}.

While most recommender systems act on data with ratings, positive-only data (e.g., click-through data, browsing history, investment history) without ratings are of interest to several physics-rooted recommendation methods. In this context, the input data are represented as a bipartite network (e.g., a user-item network~\cite{zhou2007bipartite}) or a tripartite network (e.g., a user-item-tag network~\cite{zhang2010personalized}) based on the actual links in the dataset. The recommendation can be obtained by employing classical physics processes such as diffusion (like in the Probs algorithm~\cite{zhou2007bipartite}) and heat transfer (like in the Heats algorithm~\cite{zhang2007heat}) on the network. We refer to~\cite{yu2016network} for a review of network-based recommendation algorithms.

\subsection{Recommender systems in investment domains}

Applying recommender systems to financial investment problems has received growing attention \cite{zibriczky2016recommender}. It is generally considered as a challenging task because of the strict expectations from the information-seeker~\cite{zibriczky2016recommender}.
Fields of application for recommender systems in finance-related domains include stock investment~\cite{ geva2014empirical}, real estate investment~\cite{ginevivcius2011recommender}, crowdfunding projects~\cite{an2014recommending}, and portfolio management~\cite{paranjape2013stock}, among others. As for venture investment, there have been relatively few scientific publications on the topic of recommendation. Stone et al.~\cite{stone2013empirical} used collaborative filtering to recommend relevant investment opportunities to venture capital (VC) firms. They reported that this class of activities is characterized by extremely sparse data, and the number of invested companies for VC firms follows a power-law distribution, which points out the existence of very active VC firms. Zhao et al.~\cite{zhao2015risk} proposed 5 risk-aware startup selection methods and ranking algorithms to predict VC firms' new investments.
Existing studies largely concentrated on helping VC firms to find investee companies, while there is a lack of research and effective methods to support new startups searching for investors. Bringing network-based recommendation techniques to the problem of recommending investors to startups is one of the main contributions of our paper.

\subsection{Cold-start problems}

In cold-start problems, because of insufficient past information, it is hard to infer users' preferences or items' potentially relevant users.
A simple approach to mitigate the cold-start problem is to recommend the most popular objects based on historical data. But this strategy can only provide a uniform recommendation to all users. More diverse outcomes can be obtained by gathering rapidly additional information on users' preferences. 
This can be achieved by actively eliciting the user to make more informative choices~\cite{rubens2015active}, by integrating information from other user activities~\cite{cantador2015cross}, or by using hybrid techniques to combine recommendations obtained by different methods~\cite{lika2014facing}.
As an alternative to these techniques, one can leverage the additional information to construct a network and run a standard diffusion-based algorithm on it.
Following this idea, Zhang et al.~\cite{zhang2010solving} proposed a diffusion-based recommendation algorithm which considers social tags as a bridge
connecting users and objects. They indicated that tags can effectively build up relations between existing objects and new ones, thereby providing solid recommendations of new objects. Deng et al. \cite{deng2017inferring} introduced the Social Mass Diffusion (SMD) method based on a mass diffusion process in the combined network of users' social network and user-item bipartite network. They showed that the SMD can generate more personalized recommendations for new users than the global ranking based on popularity.

Besides, scholars introduced a number of strategies based on matrix factorization algorithm.
Gantner et al.~\cite{gantner2010learning} proposed a method by an extension of matrix factorization optimized for Bayesian Personalized Ranking. They leveraged the mapping functions to compute adequate latent feature representations for new entities from their attributes.
Kula~\cite{kula2015metadata} estimated feature embeddings by factorizing the collaborative interaction matrix. In his approach, new users or items can be represented in terms of combinations of metadata features that have been estimated from the training set.
Fern{\'a}ndez-Tob{\'\i}as et al.~\cite{fernandez2016alleviating} analyzed several solutions to the new user problem in collaborative filtering based on users' personality information (using 5 factors to describe an individual's personality: openness,
conscientiousness, extraversion, agreeableness and neuroticism, as assessed by self-descriptive sentences or adjectives), including personality-based matrix factorization, personality-based active learning, and personality-based cross-domain recommendation. 

To summarize, the key element to address the cold-start problem is the acquired information about the users' preference obtained either by directly asking for auxiliary information or by actively collecting it when available in the studied platforms. But requiring the targets to provide detailed individual information is not practicable in most cases. Therefore, the majority of approaches can only mitigate the cold-start problem when a target user has at least a limited historical activity, whereas recommendations for new users without prior activity must draw support from implicit preferences or features. In this paper, we focus on the cold-start problem for startups without prior activity, and we leverage the startups' self-reported tags to infer investors' preferences. 

\section{Data and network construction}
\label{Sec:Methods}

\subsection{Dataset description}
\label{Subsec:Dataset}

We collected 45,943 investment events from \emph{www.itjuzi.com}~\cite{itjuzi}. Each event in the data includes the information of one investor, one investee company, the company's tags, and the event time (with the temporal resolution of one day). We note that joint investments that involve several investors appear as multiple events. We removed from the data all the events where the investor information is missing, which leaves us with 34,469 investment events ranging from Dec. 1$^{st}$, 1999 to Apr. 28$^{th}$, 2017. There are 5,588 investors, 14,887 companies, and 1,080 tags involved in these events. Table \ref{investment_event} shows two examples.

\begin{figure}[h]
\centering
\includegraphics[width=0.8\linewidth]{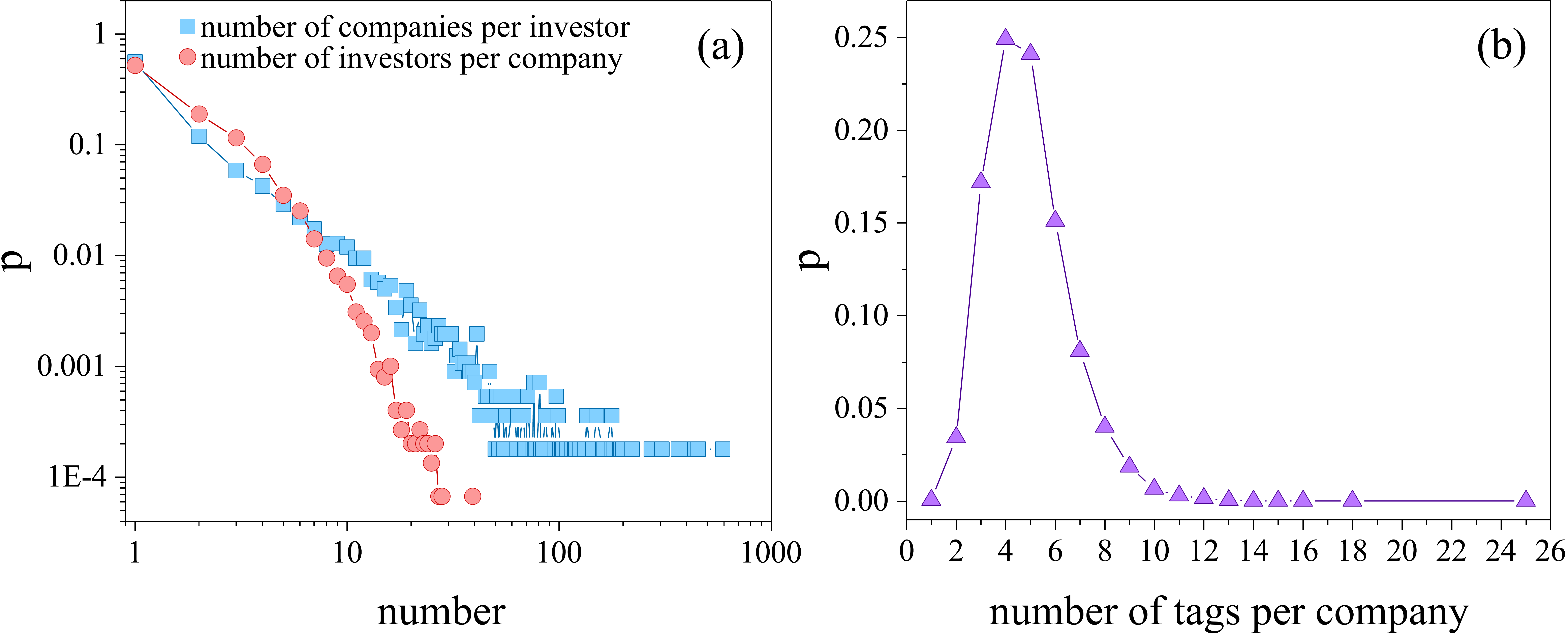}
\caption{\label{Statistical_properties}Statistical properties of the investment dataset. In panel (a), the blue squares represent the distribution of the number of companies per investor, whereas the red dots represent the distribution of the number of investors per company. Panel (b) shows the distribution of the number of tags per company.}
\end{figure}

On average, an investor invests in $6.2$ companies -- the blue squares in Figure \ref{Statistical_properties} (a) represent the distribution of the number of companies per investor. On the other hand, on average, a company has $2.3$ investors -- the red dots in Figure \ref{Statistical_properties} (a) represent the distribution of the number of investors per company. $52\%$ of companies have only \emph{one investor}, which is a manifestation of the sparse nature of investment data. Such a sparsity is reasonable since investment events are not as frequent as online activity events (such as watching movies or making new friends in online platforms)~\cite{zhao2015risk}, and the final investment decision is usually time-consuming~\cite{davila2003venture}. As for tags, each company reported 4.9 tags, on average -- Figure \ref{Statistical_properties} (b) illustrates the distribution of the number of tags per company.

\begin{table}[t]
\centering
\begin{tabular}{c c c c}
\hline
investor    &                company &  company's tags                   & time\\
\hline
\multirow{5}{*}{
IDG Capital}&\multirow{5}{*}{Tencent}& social network                    & \\	
            &                        & comprehensive social communication& \\ 
            &                        & comprehensive financial service   & 2000.04.01\\
            &                        & comprehensive game service        & \\
            &                        & SEO/SEM                           & \\
\hline
\multirow{5}{*}{
Google}     & \multirow{5}{*}{Baidu} & platform&\\ 
            &                        & enterprise service&\\
            &                        & comprehensive enterprise service  & 2004.06.01\\
            &                        & search engines                    &\\
            &                        & local comprehensive life          &\\
\hline
\end{tabular}
\caption{\label{investment_event}Two examples of investment events that involve well-known investors and companies. Notice that all the reported entries are recorded in Chinese in the original dataset.}
\end{table}

\subsection{The strong domain preference of investors}
\label{Subsec:Prefrence}

\begin{figure}[h]
\centering
\includegraphics[width = 0.75\linewidth]{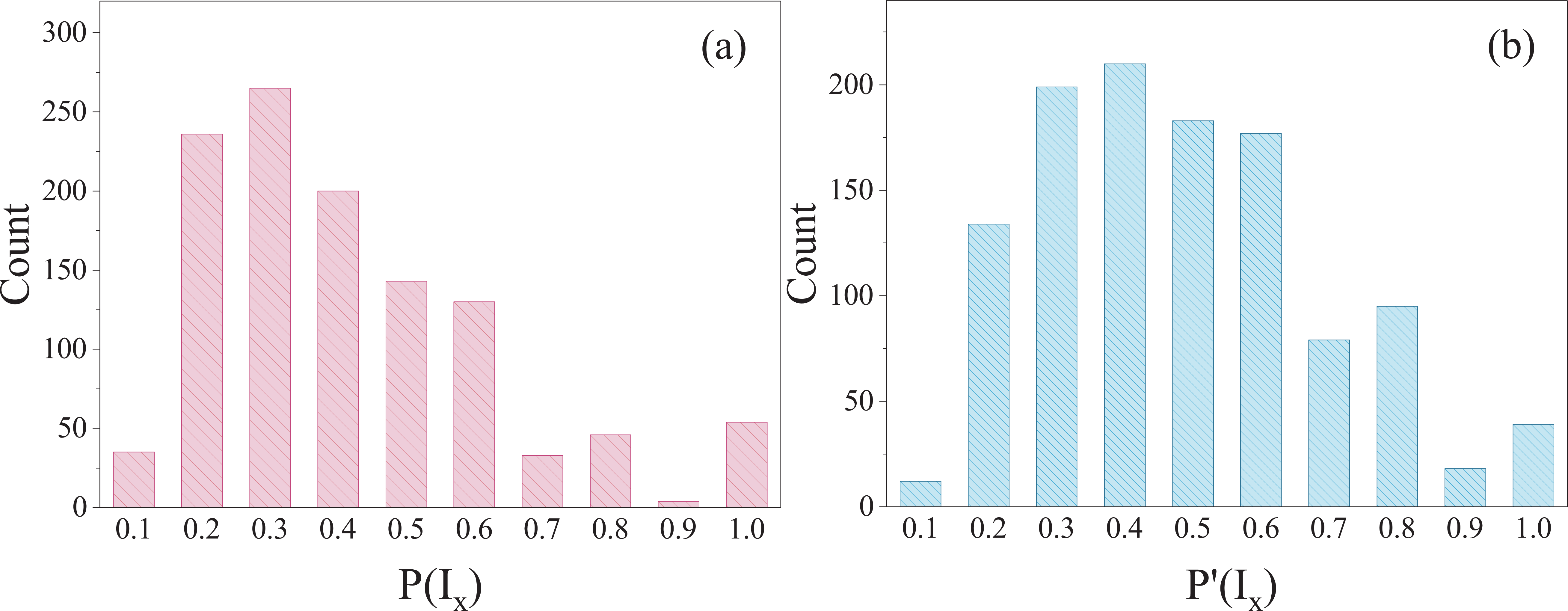}
\caption{\label{investor_preference} The strong domain preference of investors. Investors invested in less than five companies are not taken into account since their preferences are relatively difficult to capture. Panel (a) shows the distribution of the probability that one investor invests in a company with the most preferred tag of this investor. Panel (b) shows the distribution of the fraction of companies that belong to the giant component of the similarity network of the companies in which the investor has invested.}
\end{figure}

Our first main result is that investors tend to invest in specific tags (which correspond to specific domains). Evidence for the strong preference of investors toward specific tags is reported in Figure \ref{investor_preference}.
First, for each investor $I_x$, we determine her most frequent investing tag (denoted as $T_i^*(I_x)$) among those that belong to the companies that $I_x$ has invested in. 
Subsequently, we determine the fraction of $I_x$'s investments toward startups that feature tag $T_i^*(I_x)$ -- this fraction represents the strength of $I_x$'s preference toward its preferred tag, and we denote it as $P(I_x)$.
The distribution of $P(I_x)$ for all investors is shown in Figure \ref{investor_preference} (a). For each investor, on average, $45\%$ of the companies she invested in feature her preferred tag. This indicates that many investors narrow down their focus to a limited number of fields, and their preferred field dominates their investment activity. To further illustrate this property, in Figure \ref{investor_preference} (b), we assess the similarity between companies in which an investor has invested. More specifically, for each investor, we build a company similarity network where two companies are connected if they share at least one common tag.
We consider the fraction of companies that belong to the giant component of the network as the measure of company similarity for investor $I_x$, and we denote it as $P'(I_x)$. The distribution of $P'(I_x)$ for all investors is shown in Figure \ref{investor_preference} (b), and the average value is as high as 52\%. 

Overall, Figure \ref{investor_preference} reveals investors prefer to focus on a few industry categories. The reason might be that investors are cautious in investing in unfamiliar fields to avoid risk~\cite{mason2004investors}, which results in a low diversity in their investment activity. 

\subsection{Tripartite network construction}

\begin{figure}[t]
\centering
\includegraphics[width = 1\textwidth]{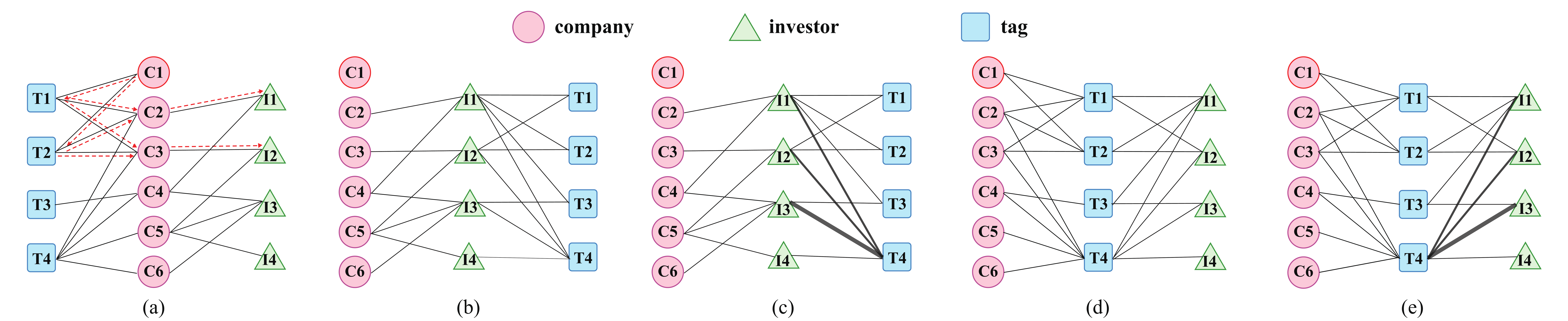}
\caption{\label{illustration_tripartite}An illustration of different possible tripartite network representations. The thickness of each line is proportional to its weight. The weight of the link between a given investor and a given tag indicates the number of companies with that tag the investor has invested in. (a) Traditional tag-company-investor (TCI) network. (b) Unweighted company-investor-tag (CIT) network. (c) Weighted company-investor-tag (CIT) network. (d) Unweighted company-tag-investor (CTI) network. (e) Weighted company-tag-investor (CTI) network.}
\end{figure}

Investment relationships can be modeled by means of an investor-company bipartite network, which represents interactions between investors and startup companies. However, in our problem of interest, the target startups have no past connections with any investors. Therefore, none of the existing recommendation techniques based on bipartite network representations are relevant to the problem.
A natural network representation for our problem is the tag-company-investor (TCI) tripartite network, which simply integrates the investor-company bipartite network and the company-tag bipartite network, as shown in Figure \ref{illustration_tripartite} (a). 
An analogous representation is commonly used in existing studies on online social networks~\cite{shang2010collaborative,zhang2010solving,zhang2010personalized}.
Differently from online systems where users can freely assign tags to their collected items, in investment systems, companies' tags are self-reported and chosen from a tag library.

Formally, we denote the TCI network representation as $\mathcal{G}(T,C,I)$, where $T,C,I$ represent the set of tags, companies and investors, respectively. In this network, link $e(I_x, C_i)$ means that investor $I_x$ invested in company $C_i$, whereas link $e(C_i, T_a)$ means that company $C_i$ has the tag $T_a$. Considering the problem where a given target startup -- $C_1$ in Fig. \ref{illustration_tripartite} -- needs a recommendation, yet it has no direct connections to investors, we are not able to find any related investors unless following the red dashed arrows to identify the investors who have invested in companies with common tags to the target company. This is also the way how tags work in recommending investors to startups through the TCI network representation. So tags act as the bridge between startups without investment and startups that already received investments. 

However, this type of tripartite network has one marked disadvantage. A popular tag which is connected to many companies, like $T_4$ in Figure \ref{illustration_tripartite} (a), will connect the target startup with many companies, and among them, unrelated companies may bring uninterested investors into consideration. Therefore, a startup that owns some popular tags, like $C_2$, will find many noisy investors following network paths. Besides, we found that many investors prefer to invest in some specific fields with certain tags (Section \ref{Subsec:Prefrence}), which cannot be leverage by the TCI representation, as tags and investors are on the opposite sides of the tripartite network.

To overcome the limitations of the TCI network representation, we construct two new tripartite networks, i.e., the company-tag-investor (CTI, $\mathcal{G}(C,T,I)$) and the company-investor-tag (CIT, $\mathcal{G}(C,I,T)$) network. These two representations include virtual connections between investors and tags. We create these virtual links between investors and tags as follows: if investor $I_x$ invested in company $C_a$, and $C_a$ has tags $T_i$ and $T_j$, then links $e(I_x, T_i)$ and $e(I_x, T_j)$ are created. Moreover, we can also construct the corresponding weighted representations by introducing weighted links: for example, weight $w_{xi}=2$ means that investor $I_x$ invested in two companies with tag $T_i$. As shown in Figure \ref{illustration_tripartite} (b)-(e), in the constructed networks, there are direct links between investors and tags which can be either weighted or unweighted. 

\begin{figure}[t]
\centering
\includegraphics[width = 0.75\textwidth]{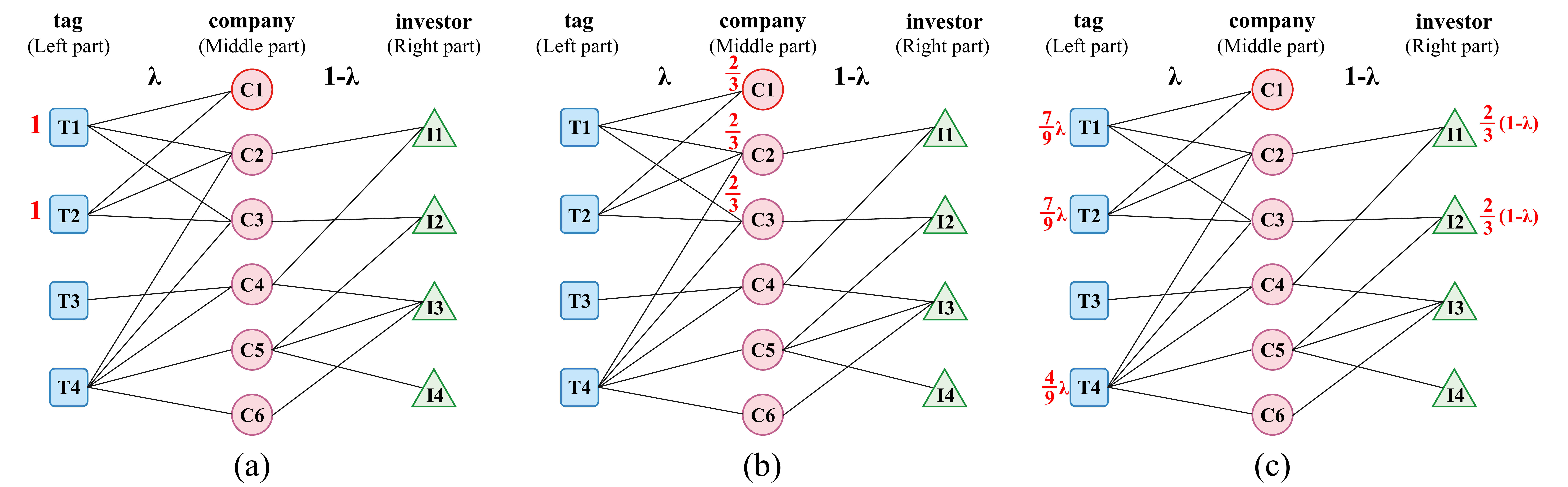}
\caption{\label{Probs}Implementing Probs algorithm on tag-company-investor tripartite network. In (a), given the target company $C_1$, we assign $C_1$'s tags, i.e., $T_1$ and $T_2$, one unit resource. In (b), all nodes redistribute their resources to their neighbors, and at the same time, all nodes update their resource values following Eq. (2). In (c), all nodes redistribute their resources to their neighbors once again.}
\end{figure}

\section{Methods}
\label{Sec:Algorithms}

\subsection{Recommendation algorithms on tripartite networks}

Based on tripartite networks, the probabilistic spreading (Probs) recommendation algorithm was first proposed for bipartite networks~\cite{zhou2007bipartite} and later proved to be effective for tripartite networks~\cite{zhang2010solving,zhang2010personalized}.
The Probs algorithm assigns an initial resource to specific nodes, then redistributes this resource in a similar way to a random-walk process. 
In the following, to generalize the representations in Fig \ref{illustration_tripartite}, we consider a generic L-M-R tripartite network representation and denote the three classes of nodes as Left nodes (L), Middle nodes (M) and Right nodes (R), respectively.
Suppose that initially, each node possesses a kind of resource, which are denoted as $f(L_x),f(M_y),f(R_z)$, namely the resource possessed by left node $L_x$, middle node $M_y$, and right node $R_z$, respectively. Subsequently, each node distributes its resource to all its neighbors. The new resource values for Left, Middle and Right nodes are determined by the equations,
\begin{eqnarray}  
\left\{  
\begin{array}{lr}  
f'(L_x)=\sum\limits_{M_y\in M}\frac{f(M_y)A(L_x,M_y)}{k_{M_y\rightarrow{L}}},\\
f'(R_z)=\sum\limits_{M_y\in M}\frac{f(M_y)A(R_z,M_y)}{k_{M_y\rightarrow{R}}},\\
f'(M_y)=
\sum\limits_{L_x\in L}\frac{f(L_x)A(L_x,M_y)}{k_{L_x}}+
\sum\limits_{R_z\in R}\frac{f(R_z)A(R_z,M_y)}{k_{R_z}},
\end{array}  
\right.
\end{eqnarray}
where \emph{A} is the adjacency matrix, $A(L_x,M_y)=1$ if $L_x$ and $M_y$ are connected. $k_{M_y\rightarrow{L}}$ is the number of neighboring left nodes for $M_y$, while $k_{M_y\rightarrow{R}}$ is the number of connected right nodes for $M_y$. $k_{L_x}$ and $k_{R_z}$ mean the degree of $L_x$ and $R_z$, respectively. Note that if the links are weighted, $k$ represents the sum of weights on the related links.

Considering that the middle nodes have two kinds of neighbors whose importance may be different, we introduce a tunable parameter $\lambda\in[0,1]$ as in~\cite{zhang2010personalized}. During the process of distribution, each middle node distributes a ratio $\lambda$ of its resources to the neighboring left nodes, and the rest $1-\lambda$ resources are distributed to the neighboring right nodes. 
We calculate the new resource through 
\begin{eqnarray}  
\left\{  
\begin{array}{lr}  
f'(L_x)=\lambda\sum\limits_{M_y\in M}\frac{f(M_y)A(L_x,M_y)}{k_{M_y\rightarrow{L}}},\\
f'(R_z)=(1-\lambda)\sum\limits_{M_y\in M}\frac{f(M_y)A(R_z,M_y)}{k_{M_y\rightarrow{R}}},\\
f'(M_y)=
\sum\limits_{L_x\in L}\frac{f(L_x)A(L_x,M_y)}{k_{L_x}}+
\sum\limits_{R_z\in R}\frac{f(R_z)A(R_z,M_y)}{k_{R_z}}.
\end{array}  
\right.
\end{eqnarray}
Given a target startup which needs recommendation and one of the tripartite network representations in Figure \ref{illustration_tripartite}, we start the iterative process described above by assigning one unit resource to all the tags of the target startup. Subsequently, the resource spreads along the network edges. A visual representation of the 2-step ProbS process is given in Figure \ref{Probs}. In this example, investors can only receive resource and update their score (new resource value) at even steps (step two, four, six, \dots). By contrast, in the company-investor-tag and company-tag-investor representations, investors can obtain new resources and update their scores at odd steps (step one, three, five, \dots). To perform a clear comparison between different tripartite networks with different diffusion steps of the algorithms, we replace \emph{step} with \emph{reach} by referring to the state of the resource after the resource arrives at the investors for the \emph{i}-th time as the $i$-reach of the process. In the TCI representation as an example, 1-reach represents the $2^{nd}$ step, 2-reach represents the $4^{th}$ step, and so forth. In the CIT representation, 1-reach represents the $1^{st}$ step, 2-reach means $3^{rd}$ step, and so forth.

For the sake of completeness, we also consider an alternative diffusion process on the tripartite network representations introduced above: the heat spreading (Heats) algorithm~\cite{zhang2007heat}, which employs a process analogous to heat diffusion. We denote by $h(L_i)$ the level of ``heat" of node $L_i$. The new level of ``heat" of each node after diffusion process are calculated following these equations,
\begin{eqnarray}  
\left\{  
\begin{array}{lr}  
h'(L_x)=\lambda\sum\limits_{M_y\in M}\frac{h(M_y)A(L_x,M_y)}{k_{L_x}},\\
h'(R_z)=(1-\lambda)\sum\limits_{M_y\in M}\frac{h(M_y)A(R_z,M_y)}{k_{R_z}},\\
h'(M_y)=
\sum\limits_{L_x\in L}\frac{h(L_x)A(L_x,M_y)}{k_{M_y\rightarrow{L}}}+
\sum\limits_{R_z\in R}\frac{h(R_z)A(R_z,M_y)}{k_{M_y\rightarrow{R}}}.
\end{array}  
\right.
\end{eqnarray}
Similar to Probs, we can get \emph{i}-reach Heats.

\begin{figure}[t]
\centering
\includegraphics[width =1\textwidth]{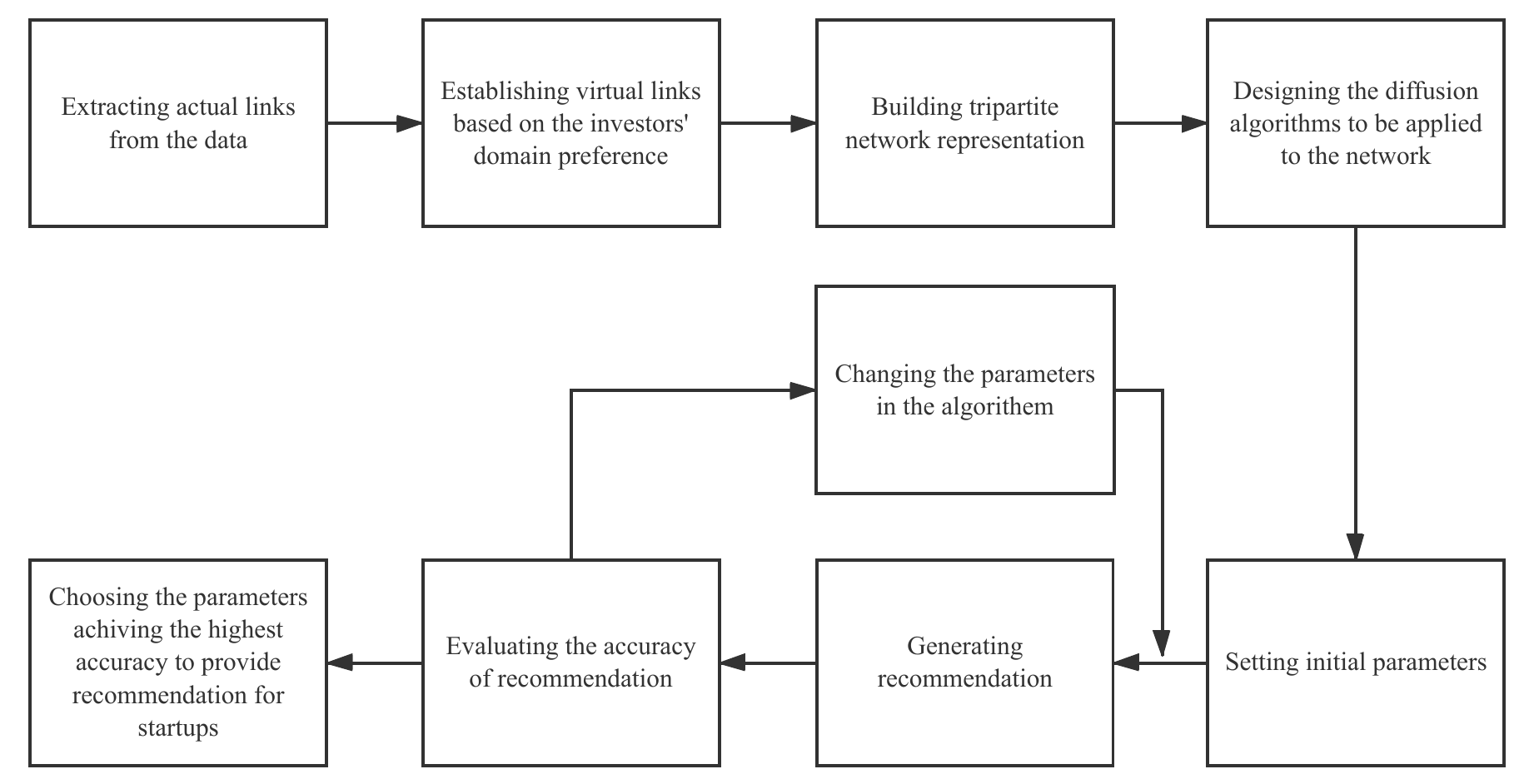}
\caption{\label{flow_chart} The working flow of the proposed recommendation process.}
\end{figure}

For Probs or Heats, each investor can obtain the value of ``resource" (or ``heat") after \emph{i}-reach diffusion process (\emph{i}=1, 2, 3, \dots). The investors with higher values will be the top choice to build the recommendation list. The working process of our approach is shown in Figure \ref{flow_chart}.

\subsection{Benchmark algorithms}
\label{Sec:Benchmark}

The techniques introduced above build on diffusion processes on tripartite network representations. Besides assessing their relative performance, it is essential to compare their performance against simpler algorithms that do not require an iterative process to compute the recommendation scores, and against state-of-the-art recommendation techniques that are potentially relevant to our recommendation problem. In total, we consider five algorithms as benchmark algorithms: three of them are simple metrics, whereas two of them are more sophisticated, state-of-the-art methods. We provide below the details of the benchmark algorithms.

\paragraph{Popularity-based metric} 

We measure the number of investments that each investor has previously made -- i.e., the investors' degree in the investor-company bipartite network as the recommendation score. The more investments an investor has previously made, the higher its ranking position in the recommendation list for new startups that did not yet receive any investment.

\paragraph{Tag-voting metric}

We consider simple metrics that leverage tag information. For a target startup $C_a$ with tags$\{T_1,\dots,T_k\}$, if investor $I_x$ invested in $n_i$ companies with tag $T_i$, the score of $I_x$ for $C_a$ is defined as
\begin{equation}
s(I_x, C_a)=\sum\limits_{i=1}^k n_i
\end{equation}
We refer to the recommendation of investors based on this score (the investors with the largest score are recommended to the target startup) as the tag-voting method.

\paragraph{Normalized tag-voting metric}

To prevent the potential bias toward popular tags of the tag-voting metric, we further consider a normalized metric defined as
\begin{equation}
s(I_x,C_a)=\sum\limits_{i=1}^k\frac{n_i}{N_i}
\end{equation}
where $N_i$ is the total number of investments received by companies with tag $T_i$. We refer to the recommendation based on this metric (again, the investors with the largest score are recommended to the target startup) as the normalized tag-voting method.

\paragraph{Neighbor-based Collaborative filtering}

Collaborative filtering technology is by far one of the most widely used and successful method~\cite{isinkaye2015recommendation}. A large number of studies has proved its high accuracy in various systems, such as E-commerce platforms~\cite{linden2003amazon}, digital research libraries~\cite{torres2004enhancing}, online social systems~\cite{yang2014survey}, and so on.
Here we also tested an extend collaborative filtering method as a benchmark. Using tag information, one can measure the similarity between two companies, and leverage the investment history of similar companies to generate a recommendation. According to ~\cite{sarwar2001item}, the score of investor $I_x$ for startup $C_a$ is defined as 
\begin{equation}
s(I_x, C_a)= \frac{\sum\limits_{b}^{N_c}sim(C_a,C_b)\,A(I_x,C_b)}{\sum\limits_{b}^{N_c}sim(C_a,C_b)}
\end{equation}
where $N_c$ is the number of companies in the dataset, $sim(C_a,C_b)$ is the similarity between company $C_a$ and $C_b$, $A(I_x,C_b)=1$ if investor $I_x$ has invested in company $C_b$. Following~\cite{sarwar2001item}, in this paper, we leverage the cosine similarity~\cite{sarwar2001item} which is given by
\begin{equation}
sim(C_a,C_b)= \frac{T(a)\cap T(b)}{\sqrt{|T(a)|\times |T(b)|}}
\label{sim}
\end{equation}
where T(a) refer to the tags belong to company $C_a$, $|T(a)|$ is the size of $T(a)$, $T(a)\cap T(b)$ is the common tags for $C_a$ and $C_b$. We name this metric as neighbor-based CF.

\paragraph{Matrix factorization based on Bayesian Personalized Ranking}

Matrix factorization with Bayesian Personalized Ranking learning (BPR) framework has been applied to various recommender systems and generally considered as a powerful recommendation method in implicit or positive-only feedback dataset~\cite{cunha2016selecting,hong2013co}, which fit the characteristic of our collected data -- investment behavior is in “one-class” (investing) form.
We consider the combination of the matrix factorization based on BPR and feature mapping. Following Gantner et al.'s work~\cite{gantner2010learning}, we used a matrix factorization model based on the BPR framework to train the latent-feature factor of investors and companies. Then the latent-feature vector of a given new startup can be approximated based on those of similar companies. With this estimation, we can compute a score for the new startup $C_a$ with an investor $I_x$ through
\begin{equation}
s(I_x, C_a)=\left<W(I_x), \phi(C_a)\right>
\end{equation}
where $W(I_x)$ is the trained factor for $I_x$, $\phi(C_a)$ is $C_a$'s estimated latent factor.
The similarity between companies is determined through Equation~\eqref{sim}. And we employ the weighted
$k$-nearest-neighbor (kNN) regression~\cite{trevor2009elements} to estimate the factor of the new startup. For example, $C_a$'s latent factor is estimated through the equation as in ~\cite{gantner2010learning}
\begin{equation}
\phi(C_a)= \frac{\sum\limits_{b \in N_k(C_a)}sim(C_a,C_b)\,\phi(C_b)}{\sum\limits_{b \in N_k(C_a)}sim(C_a,C_b)}
\end{equation}
where $N_k(C_a)$ represents the set that comprises the $k$ most similar companies to $C_a$ and $k$ is set to 6 as it is a suitable value on the training set.
The factor number is equal to 30 using
hyperparameters that typically yields satisfactory results in non-cold start evaluations in our dataset. We also tested different numbers of factors (10, 20, 30, 40, 50, 60), and results are qualitatively similar. We refer to this method as BPR-MF method.
\subsection{Evaluation}

To evaluate the proposed recommendation methods, we divide the dataset into two parts, the training set and the testing set. The training set includes 31,039 investment events that happened before Sept. 6, 2016 (i.e., the 90\% earliest events in the data). Based on it, we construct the different tripartite network representations described above. The remaining 3,430 investment events (i.e., the 10\% latest events in the dataset) compose the testing set. For the validation of information-filtering predictive techniques, there is no unique or universal criterion to choose the relative size or temporal duration of training and testing set. The $90\%/10\%$ proportion adopted here a standard choice in the evaluation of recommender systems. Yet, we also tested different sizes ($85\%/15\%$ and $95\%/5\%$) -- the obtained results are in qualitative agreement with those obtained with the $90\%/10\%$ splitting, and they are discussed in Appendix~\ref{otherResults_comparison} and \ref{moreResults_comparison}. We stress that our aim is to recommend investors to new startups; as a consequence, only the $1,096$ companies that obtained their first investment within the testing set and their corresponding investors are considered in the evaluation -- we refer to them as \emph{target startups}. 

\emph{Ranking Score} (\emph{RS})~\cite{zhou2007bipartite} is used to evaluate the various recommendation methods. 
For a target startup, we refer to the investors who actually invested in the target startup within the testing set as the \emph{relevant investors} to that startup and refer to all the other investors as \emph{irrelevant investors}.
The \emph{RS} measures the relative ranking of the relevant investors in the target startups' recommendation list: when there are \emph{o} investors that can potentially be recommended, a relevant investor with ranking \emph{r} achieves the relative ranking \emph{r/o}, which is equal to her ranking score. By averaging over all target startups in the testing set and their relevant investors, we obtain the mean \emph{RS}: the smaller the mean ranking score, the higher the algorithm's accuracy. We stress that to evaluate the recommendation methods' performance in the cold-start problem studied here, the ranking score is a better evaluation metric than \emph{Precision} and \emph{Recall} because the target startups typically have only few investors.

We also measure the \emph{AUC} (Area under receiver operator curve)~\cite{Hanley1982meaning} as an alternative performance metric. The AUC aims at assessing the method's ability to distinguish relevant from irrelevant investors. For a target company, its \emph{AUC} can be approximated as the probability that, when choosing at random one investor from the relevant investors and another investor from the irrelevant investors, and the relevant investor's score is higher than the irrelevant investor's score. We get the mean \emph{AUC} by averaging over all target companies in the test set; the higher the AUC, the better the method's performance.

\section{Results}

\label{Sec:Results}
\begin{figure}[t]
\centering
\includegraphics[width =1\textwidth]{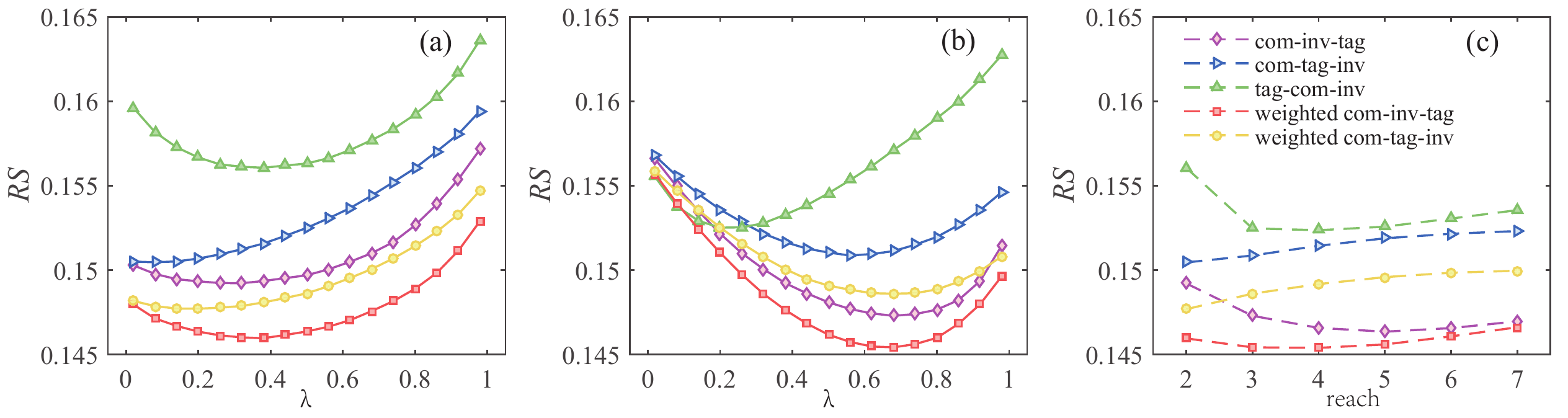}
\caption{\label{Results_comparison} The comparison of Probs recommendation algorithms applied to different tripartite networks. Each line corresponds to a specific network representation. (a) and (b) show the average \emph{RS} values for 2-reach Probs and 3-reach Probs, respectively, as a function of the algorithm parameter $\lambda$. (c) shows
the average RS value attained by the Probs algorithm with optimal $\lambda$, as a function of the reach of the diffusion process. Note that in the legends, com and inv stand for company and investor, respectively.}
\end{figure}

Figure~\ref{Results_comparison} compares the recommendation results for the Probs algorithm for different network representations. Figure \ref{Results_comparison} (a) and (b) show the \emph{RS} values for 2-reach Probs diffusion and 3-reach Probs diffusion, respectively, as a function of $\lambda$, which represents the proportion of resources distributed from central to left and right nodes. Generally, the proposed tripartite network representations that leverage virtual links between investors and tags ((i.e., company-investor-tag and company-tag-investor tripartite networks) outperform the natural tag-company-investor representation. Besides, algorithms based on weighted network representations outperform those based on unweighted representations (see Figure \ref{Results_comparison}).

Comparing Figures \ref{Results_comparison} (a) and (b), we find that the average \emph{RS} depends on the reach of the considered diffusion process. To identify the best overall method, we obtain the best \emph{RS} values for all the tripartite networks at 2-reach diffusion, 3-reach diffusion, $\cdots$, 7-reach diffusion respectively, and show them in Figure \ref{Results_comparison} (c). We conclude that both the optimal \emph{RS} curve (see Figure \ref{Results_comparison} (c)) and the optimal \emph{RS} value (see Table \ref{Rankingscore}) are obtained from the weighted company-investor-tag network.

\begin{table}[htbp]
\centering
  \begin{tabular}{c|cc|cc|c}
  \hline
  \hline
  \multirow{3}{*}{\textbf{Network}}&\multicolumn{2}{c|}{ \textbf{Com-inv-tag}}& \multicolumn{2}{c|}{\textbf{Com-tag-inv}}& \textbf{Tag-com-inv}\\
  \cline{2-6}
  & \textbf{weighted}& \textbf{unweighted} &\textbf{weighted}& \textbf{unweighted} & \multirow{2}{*}{\textbf{unweighted}}\\
  & \textbf{inv-tag}&\textbf{inv-tag}&\textbf{inv-tag}&\textbf{inv-tag}\\
  \hline
  Probs 1-reach& 0.21094&0.22334& 0.21094&0.22334& 0.21478 \\
  \hline
  \multirow{2}{*}{Probs 2-reach}&0.14596&0.14922&0.14769&0.15046&0.15607\\
  & $\lambda$*=0.36 &$\lambda$*=0.28& $\lambda$*=0.18 &$\lambda$*=0.10&$\lambda$*=0.38 \\
  \hline
  \multirow{2}{*}{Probs 3-reach}&0.14543&0.14730&0.14860&0.15087&0.15248\\
  &$\lambda$*=0.68&$\lambda$*=0.68&$\lambda$*=0.66&$\lambda$*=0.56&$\lambda$*=0.22\\
  \hline
  &\textbf{0.14540}&0.14636&0.14769 &0.15046&0.15237\\
  Probs reach* &\textbf{$\lambda$*=0.82}&$\lambda$*=0.92&$\lambda$*=0.18&$\lambda$*=0.10&$\lambda$*=0.14\\
  &\textbf{reach*=4}&reach*=5&reach*=2&reach*=2&reach*=4\\
  \hline
  &0.25612&0.27259&0.27073&0.27880&0.25631  \\
  Heats reach*&$\lambda$*=0.56&$\lambda$*=0.44&$\lambda$*=0.34&$\lambda$*=0.32&$\lambda$*=0.60\\
  &reach*=2&reach*=2&reach*=2&reach*=2&reach*=2\\
  \hline
  \multirow{4}{*}{benchmarks}&\multicolumn{5}{c}{popularity-based method=0.16507}\\
  &\multicolumn{5}{c}{tag-voting = 0.20781}\\
  &\multicolumn{5}{c}{normalized tag-voting = 0.21094}\\
  &\multicolumn{5}{c}{neighbor-based CF = 0.20791}\\
  &\multicolumn{5}{c}{BPR-MF = 0.35971}\\
  \hline
  \hline
  \end{tabular}
\caption{\label{Rankingscore} Average \emph{RS} of the considered recommendation algorithms for three tripartite network representations. For each value of reach, we provide the optimal $\lambda$ value together with the corresponding average \emph{RS}. We also provide the performance by the Heats algorithm and five benchmark metrics. For each network representation, the ProbS algorithm substantially outperforms the Heats algorithm. It also outperforms the benchmark metrics.}
\end{table}

\begin{figure}[t]
\centering
\includegraphics[width = 0.98\textwidth]{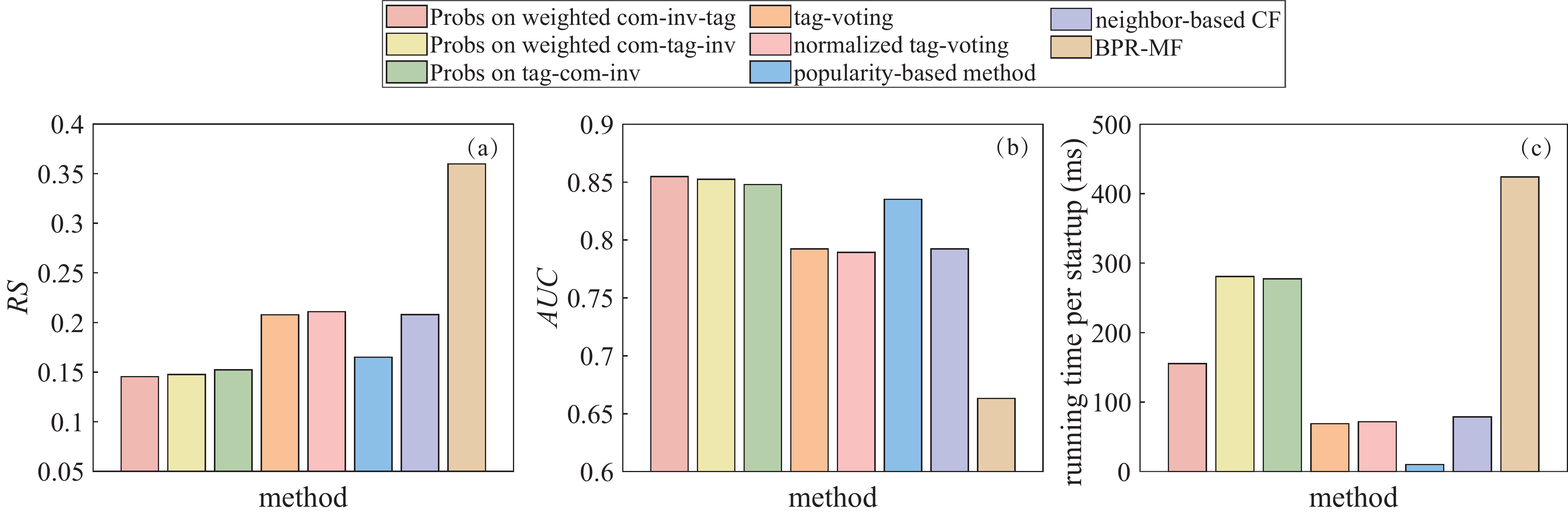}
\caption{\label{AUC_and_time}(a) Average \emph{RS} of the Probs algorithm for three tripartite network representations and five benchmark methods. The applied $\lambda$ and reach for the Probs algorithm are those achieve the optimal performance as shown in Table \ref{Rankingscore}. The smaller the \emph{RS}, the better the method's performance. (b) Average \emph{AUC} of the Probs algorithm for three tripartite network representations and five benchmark methods. Noted that $\lambda$ and reach are chosen based on the optimal performance in the \emph{RS} evaluation. The higher the \emph{AUC}, the
better the method's performance. (c) The test times per startup for different methods. We test 2-reach Probs on the three network representations.}
\end{figure}

It is essential to compare the performance of the Probs method with both the Heats algorithm and the benchmark algorithms introduced in Section \ref{Sec:Benchmark}. We find that the optimal performances achieved by the Heats algorithm (i.e., the performance achieved by selecting the optimal reach of the algorithm, for each network representation) are substantially worse than those achieved by non-optimized Probs algorithm (see Table \ref{Rankingscore}). For instance, the average \emph{RS} of Heats algorithm achieves 0.25612 with 2-reach on the weighted company-investor-tag tripartite network, as compared to 0.14596 achieved by Probs algorithm with 2-reach on the same tripartite network. Similarly, the average \emph{RS} of the three simple benchmark metrics introduced above (popularity-based, tag-voting, and normalized tag-voting metric) is larger than that obtained by the Probs algorithm with more than one reach (see Table \ref{Rankingscore} and Figure \ref{AUC_and_time} (a)). For example, among these local metrics, the best-performing one (popularity-based method) achieves average \emph{RS} equal to 0.16507, as opposed to 0.14596 achieved by Probs with two reach based on the weighted company-investor-tag tripartite network. 
As for the common-used collaborative filtering and advanced matrix factorization model, their performance is also inferior to that of the Probs algorithm on tripartite networks. This could be due to the sparsity of the investment data that may limit their effectiveness~\cite{stromqvist2018matrix}. And the results also indicate that the tag information is more powerful when it is connected to the investors' preference than when it is used to measure the similarity between companies.

To further verify the stability of the results, we repeat the predictive analysis for different proportions of the training and testing set sizes ($85\%/15\%$ and $95\%/5\%$), and for another evaluation metric (\emph{AUC}).
It turns out that the relative performance of the methods is in agreement with the findings for the $90\%/10\%$ splitting (see Figure \ref{otherResults_comparison} and \ref{moreResults_comparison}). The \emph{AUC} results as shown in Figure \ref{AUC_and_time} (b) also show the advantage of the accuracy of Probs on weighted com-inv-tag network representation, which are in agreement with the \emph{RS} results.

We conclude that: (i) the Probs algorithm performs significantly better than the Heats algorithm in the recommendation of investors to new startups; (ii) the benchmark metrics cannot compete with the Probs algorithm on the tripartite network representations, which suggests that constructing the tripartite networks considered here are indispensable for an accurate recommendation. (iii) the Probs algorithm on the weighted com-inv-tag achieves the best recommendation performance.

As for run-time overhead (Figure \ref{AUC_and_time} (c)), one can clearly see that 2-reach Probs in the weighted com-inv-tag network has a relatively fast testing time, while the other two network representations take longer times to generate the predictions. Overall, the computational time of all the considered metric is of the same order of magnitude.

\section{Conclusion and discussion}
\label{Sec:Conclusions}
The main contribution of this paper is to solve the investor recommendation problem by building tripartite network representations including virtual links in combination with the Probs diffusion algorithm. The new application scenario and the introduction of virtual links into diffusion-based recommendation are the main innovative features of our study.
Obtaining financial support is a rigid demand for most startups; thus, to precisely recommend investors for new startups is valuable, yet difficult due to the cold-start problem. We started by constructing a natural tag-company-investor tripartite network representation. In consideration of its disadvantages and the significant role played by tag in investment decisions, we reconstructed two tripartite network representations by establishing virtual connections between investors and tags, which take full advantage of the investors' preference for tags. The obtained results demonstrate that our modification of the traditional tripartite network is effective to provide better recommendations, and the weighted company-investor-tag network achieves the best performance in accuracy. Besides, the Probs algorithm significantly outperforms the Heats algorithm and baseline methods.

Recommendation in financial investment domains -- especially in the venture investing domain -- is a new topic in the recommender systems literature. A few academic studies have worked on applying recommendation techniques to provide personalized recommendations for investors -- whether individual investors or investment institutions -- but neglected the startups' demand. Our research paves the way to the extensive investigation of the investor recommendation problem for startups. 

We identify three main directions for future improvements. First, we have not analyzed the temporal effects of related data. A fundamental assumption of our method is that investors tend to invest in companies with tags which are similar to those of their past investments. But in reality, the preference of investors can change over time. Incorporating the temporal information may improve the recommendation performance. Second, it is essential to test our new tripartite networks along with other recommendation methods on more investment datasets. These attempts will provide us a promising way to better inform investment decision making and, as a result, assist startups effectively. Third, we leveraged domain (tag) information to generate effective recommendations. Incorporating into the recommendation algorithm additional factors such as financial consideration, product and market characteristics, social relation information~\cite{gompers2016venture,martinez2015model,mason2004investors}, might further improve the predictive performance.

\section*{Acknowledgements}
This work is supported by the National Natural Science Foundation of China (Grants Nos. 11622538, 61673150, 61703074), the Zhejiang Provincial Natural Science Foundation of China (Grant No. LR16A050001), and the Science Strength Promotion Programme of UESTC. MSM also acknowledges support from the University of Zurich through the URPP Social Networks, the Swiss National Science Foundation Grant No. 200021-182659, the UESTC professor research start-up grant No. ZYGX2018KYQD215.

\section*{Author contributions statement}
Q.Z. and L.L. conceived the idea and designed the research. S.X. and Q.Z. collected and processed the data. S.X. performed the experiments. All authors analyzed and discussed the results. S.X., Q.Z. and M.S.M. wrote the manuscript. All authors reviewed the manuscript.

\bibliography{Recommending_investors_for_new_startups-proof}

\appendix

\section{Different training/testing set splitting}

\begin{figure}[h]
\centering
\includegraphics[scale=0.6]{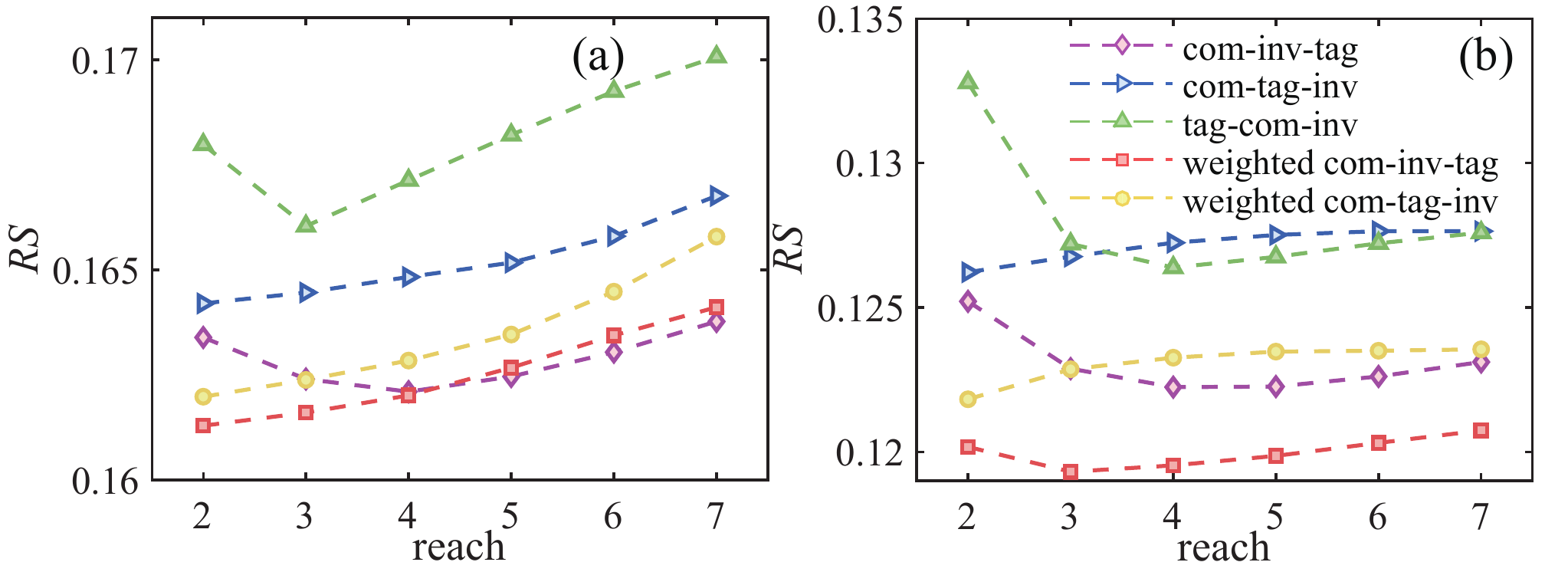}
\caption{\label{otherResults_comparison}Performance of the Probs algorithm applied to five different tripartite network representations, for different sizes of the testing set. In (a), the proportion of links in the testing set is 15\%. In (b), the proportion of links in the testing set is 5\%. The methods show a similar relative performance as in Figure \ref{Results_comparison} (c)}
\end{figure}

\begin{figure}[h]
\centering
\includegraphics[scale=0.45]{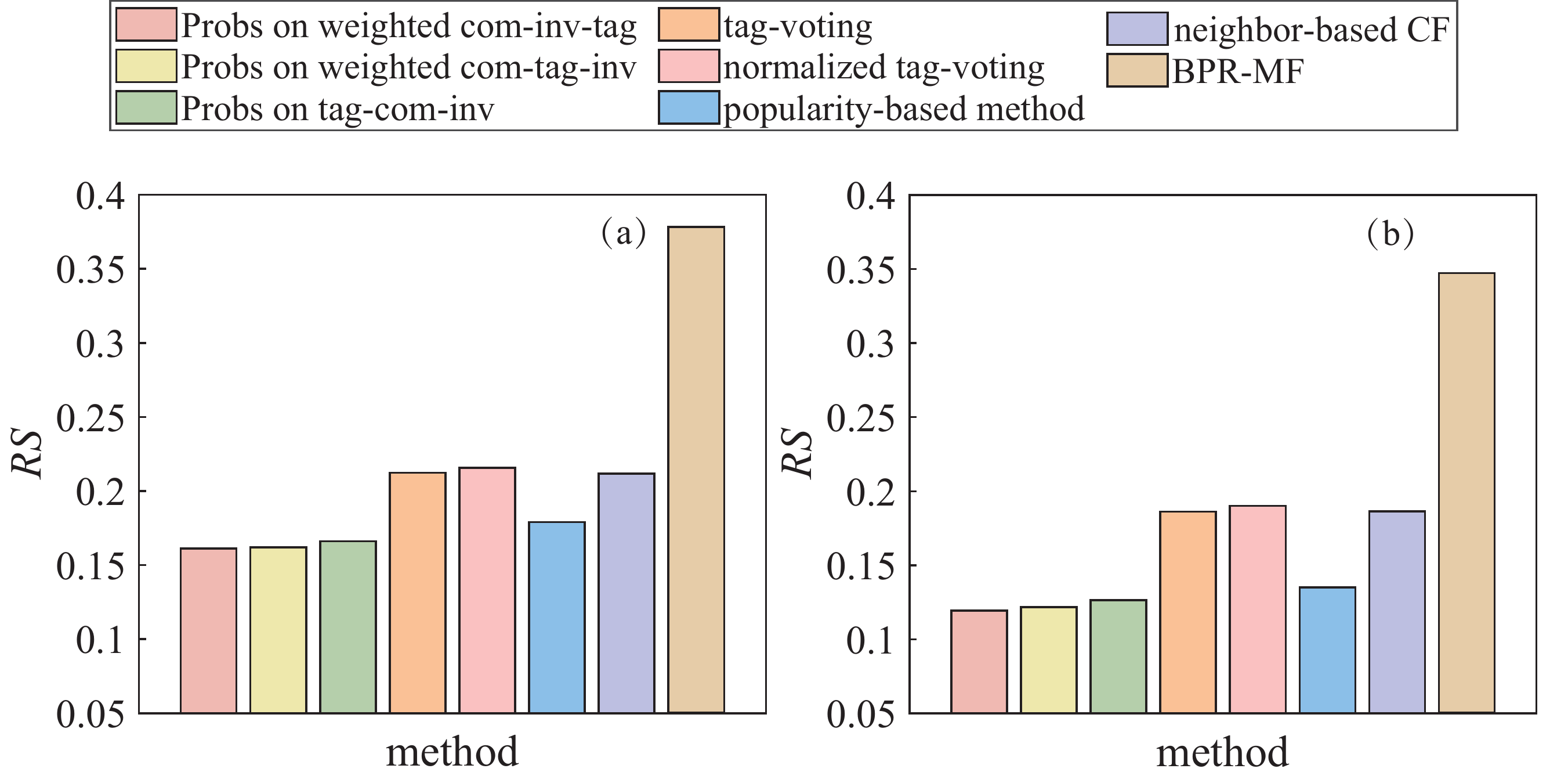}
\caption{\label{moreResults_comparison}A comparison of considered recommendation methods with different size of the testing set. In (a), the proportion of links in the testing set is 15\%. In (b), the proportion of links in the testing set is 5\%. The methods show a similar relative performance as in Figure \ref{AUC_and_time} (a).}
\end{figure}

We test the 85\%/15\% and 95\%/5\% splitting of training and testing set to examine the stability of the results. Figure \ref{otherResults_comparison} shows the similar relative performance, which is in agreement with Figure \ref{Results_comparison} (c). Figure \ref{moreResults_comparison} is in agreement with Figure \ref{AUC_and_time} (a).

\end{document}